\documentclass[prl,showpacs,twocolumn]{revtex4}

\usepackage{amssymb,bm,color,graphics,graphicx,amsfonts,amsmath,empheq}
    
\usepackage{pdfpages}

\def\d{\mathrm{d}}
\def\e{\mathrm{e}}
\def\ds{\partial_s}

\def\dt{\partial_t}
\def\um{\mu{\rm m}}
\def\pN{{\rm pN}}

\begin{document}
\title{Beat regulation in twisted axonemes}
\author{Pablo Sartor$^1$, Veikko F Geyer$^2$, Jonathon Howard$^2$ and Frank J\"ulicher$^1$}

\affiliation{$^1$Max Planck Institute for the Physics of Complex
  Systems. Noethnitzer Strasse 38 , 01187, Dresden,
  Germany. $^2$Department of Molecular Biophysics and Biochemistry, Yale University, New Haven, Connecticut}


\begin{abstract}
Cilia and flagella are hairlike organelles that propel cells through fluid. The active motion of the axoneme, the motile structure inside cilia and flagella, is powered by molecular motors of the dynein family. These motors generate forces and torques that slide and bend the microtubule doublets within the axoneme. To create regular waveforms the activities of the dyneins must be coordinated. It is thought that coordination is mediated by stresses due to radial, transverse, or sliding deformations, that build up within the moving axoneme. However, which particular component of the stress regulates the motors to produce the observed flagellar waveforms remains an open question. To address this question, we describe the axoneme as a three-dimensional bundle of filaments and characterize its mechanics. We show that regulation of the motors by radial and transverse stresses can lead to a coordinated flagellar motion only in the presence of twist. By comparison, regulation by shear stress is possible without twist. We calculate emergent beating patterns in twisted axonemes resulting from regulation by transverse stresses. The waveforms are similar to those observed in  flagella of {\it Chlamydomonas} and sperm. Due to the twist, the waveform has non-planar components, which result in swimming trajectories such as twisted ribbons and helices that agree with observations.
\end{abstract}

\maketitle

Cilia and flagella are slender cell appending organelles that contain a motile internal structure called the axoneme. The axoneme, in turn, contains nine microtubule doublets, a central pair of microtubules, motor proteins in the axonemal dynein family, and a large number of additional structural proteins  \cite{pazour_proteomic_2005,pigino2012axonemal}, see Fig.~\ref{fig:geo}A. The axoneme undergoes regular oscillatory bending waves that propel cells through fluids and fluids along the surfaces of cells. This beat is powered by the dyneins, which generate sliding forces between adjacent doublets \cite{brokaw_direct_1989}. Bending originates from an imbalance of antagonistic motors acting on doublets on opposing sides of the beating plane, see Fig.~\ref{fig:geo}A \cite{satir1989splitting, brokaw2009thinking}. 
 

It has been suggested that the switching of dynein activity on opposite sides of the axoneme is the result of feedback and non-linearities \cite{satir1989splitting, brokaw_molecular_1975, riedelkruse_how_2007, camalet_generic_2000}. The axonemal dyneins generate forces deforming the axoneme; the deformations or the corresponding stresses, in turn, regulate the dyneins. Several components of the stress can be distinguished: radial stress, that tends to increase the axoneme radius;  transverse stress, that tends to separate the doublets; and  shear stress, that tends to slide doublets with respect to each other. How these components can regulate dynein activity remains poorly understood. One hypothesis, referred to as ``geometric clutch'',  is that dynein is regulated by transverse stress.  Due to the circular symmetry of the transverse stress, motors on opposite sides of the axoneme experience the same transverse stress, so transverse stresses are not antagonistic and cannot be used as a signal for switching. As a consequence, regulation by transverse stress requires an additional asymmetry \cite{lindemann1994model, brokaw2009thinking,bayly2014equations}, whose origin remains elusive. Other suggested control mechanisms, such as regulation by sliding \cite{brokaw_molecular_1975, julicher_spontaneous_1997, brokaw2005computer,riedelkruse_how_2007} or curvature \cite{brokaw_bend_1971, brokaw2009thinking, machin_1958}, do not have this circular symmetry and thus do not require an additional asymmetry.

In addition to sliding forces, axonemal dyneins can also generate torques, which rotate microtubules in {\it in vivo} assays \cite{yamaguchi2015torque, vale1988rotation}. In the axoneme, this rotation will lead to twist \cite{hines1985contribution}. Because of the possibility that twist might be important for beat generation, we have developed a three-dimensional model of the axoneme. This model, inspired by earlier work \cite{hilfinger2008chirality}, is able to distinguish between transverse and radial stresses. We show that twist is capable of breaking the circular symmetry of these stresses. In this case, transverse stress can be used as a signal for switching and so can act as a control mechanism. The emergent beating patterns are similar to those of {\it Chlamydomonas} axonemes; they are non-planar and result in complex swimming trajectories such as twisted ribbons and helices.

\section{Continuum mechanics of the axoneme}
\begin{figure}[t]
\centering
\includegraphics{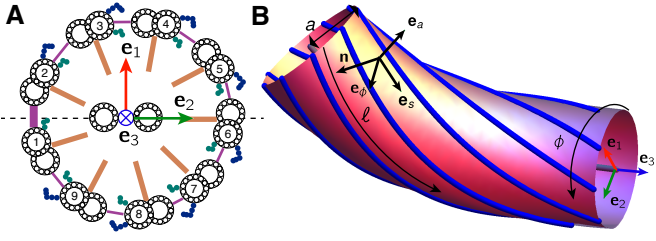}
\caption{{\bf Geometry of the axoneme.}  {\bf A}. Schematic of an axonemal cross-section, with  numbering for the $9$ doublets, as seen from the basal endß. Dyneins appear in blue and green, elastic linkers in purple, and radial spokes in orange. The cross-bridges between doublets 1-2, thick purple lines, define the beating plane, dashed line.  The center-line triad is oriented such that ${\bf e}_1$, in red,  points between doublets 3 and 4, and the beating plane is spanned by ${\bf e}_2$, in green, and ${\bf e}_3$, in blue. Bending in the direction of  ${\bf e}_2$ requires the motors in filaments $2-5$ to be active, while bending in the direction of $-{\bf e}_2$ requires activity of motors in filaments $6-9$. {\bf B}. The bundle of filaments is parametrized by the azimuthal coordinate $\phi$ and the arc-length $\ell$ which depends on the center-line arc-length $s$. These define tangent vectors ${\bf e}_s$ and ${\bf e}_\phi$. The unit vector ${\bf n}$ is normal to ${\bf e}_s$, and ${\bf e}_a$ is orthogonal to the tangent plane. The basal end is on the right. \label{fig:geo}}
\end{figure}

The axoneme contains a regular arrangement of microtubule doublets in a cylindrical geometry that is stabilized by additional structural elements such as radial spokes, a central pair of microtubules, dynein molecular motors, and other elements such as nexin linkers \cite{pigino2012axonemal}. The doublets $n=1$ and $n=2$ can be  identified by a higher density in electron microscopy images referred to as cross-bridge, see Fig.~\ref{fig:geo}A.

We characterize the axonemal structure by a bundle of filaments corresponding to the microtubule doublets that are arranged on a cylindrical sheet of radius $a$ \cite{hilfinger2008chirality}, see Fig.\ref{fig:geo}. The cylindrical sheet ${\bf R}(s,\phi)$ is parametrized by an angular coordinate $\phi$ and the distance variable $s$ as
\begin{align}
{\bf R}(s,\phi) &= {\bf r}(s)+{\bf e}_1(s)a(s,\phi)\cos(\varphi(s,\phi))\nonumber\\
&+{\bf e}_2(s)a(s,\phi)\sin(\varphi(s,\phi))\quad.
\label{eq:sheet}
\end{align}
Here, $s$ is the arc-length of the center-line ${\bf r}(s)$ of the cylindrical sheet measured from base to tip. The vectors ${\bf e}_{1}$ and ${\bf e}_{2}$ are unit vectors normal to the centerline with unit tangent ${\bf e}_3=\ds{{\bf r}}$. We choose ${\bf e}_1$ to point between filaments $n=3$ and $n=4$, and orient ${\bf e}_2$ perpendicular to both ${\bf e}_1$ and ${\bf e}_3$, see Fig.~\ref{fig:geo}A. The angular parameter $\phi$ is used to identify filaments which correspond to the values $\phi=\phi_n$ with $\phi_n=2\pi (n-1/2)/9$. The function $\varphi(s,\phi)$ describing the {azimuthal angle} of filaments indexed by $\phi$. Similarly we allow the cylinder radius $a(s,\phi)$ to depend on $s$ and $\phi$.

To characterize the geometry of the filaments  we introduce the tangent vectors ${ {\bf e}_s =\partial_s{\bf R}/|\partial_s{\bf R}|}$ and ${\bf e}_\phi =\partial_\phi{\bf R}/|\partial_\phi{\bf R}|$, which form a basis of the tangent space on the cylindrical sheet. We also introduce the filament normal ${\bf n}$ in the tangent plane. It obeys ${\bf e}_s\cdotp{\bf n}=0$ and ${\bf n}^2=1$, see Fig.~\ref{fig:geo}B. The vector ${\bf e}_a={\bf n}\times{\bf e}_s$ is normal to the cylinder pointing outwards. The filament curvatures tangential and perpendicular to the cylindrical surface are then given by  $C_1(s,\phi)={\bf e}_s\cdotp\partial_s {\bf n}$ and $C_2(s,\phi)={\bf e}_a\cdotp\partial_s {\bf e}_s$, respectively.

\begin{figure}[t]
\centering
\includegraphics{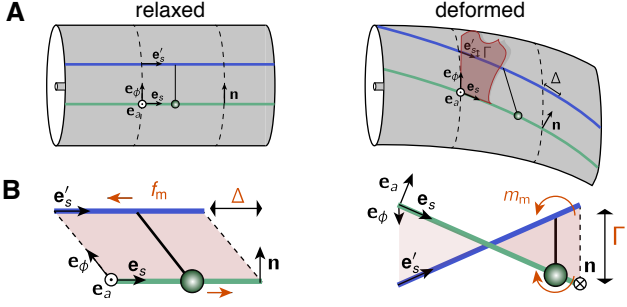}
\caption{{\bf Deformation of the axoneme}  {\bf A}. Relaxed and deformed geometries of the axoneme with two filaments marked in green and blue, and a dynein represented as two green circles. In the deformed state the tangent plane (in red) spanned by ${\bf e}_s$ and ${\bf e}_\phi$ does not contain the tangent vector ${\bf e}_s'$ of the contiguous doublet, which is displaced out of the plane by a distance $\Gamma$ in the direction of ${\bf e}_a$. The sliding $\Delta$ corresponds to the length mismatch of the two successive doublets measured by projecting in the direction of the normal vector ${\bf n}$. {\bf B}. A dynein motor creates a pair of forces $f_{\rm m}$ that produces sliding $\Delta$, and also torques $m_{\rm m}$ that produce out of plane displacement $\Gamma$.  The deformed states  are shown from different perspectives.\label{fig:rates}}
\end{figure}

In order to discuss the mechanics of the axoneme we introduce the relevant deformation variables. For simplicity, we impose the constraints $\varphi=\phi$ and $a=a_0$ for which the cylinder radius $a_0$ and the separation between neighboring filaments $2\pi a_0 /9$ are fixed. We will also impose incompressibility of the filaments, see 
\cite{heussinger2010statics} for a more general treatment. Two key deformation variables of the filament bundle are the filament sliding displacement $\Delta$ \cite{everaers1995fluctuations, hilfinger2008chirality,heussinger2010statics} and the filament splay $\Gamma$, see Fig.~\ref{fig:rates}A. The sliding displacement  is defined by 
\begin{align}
\Delta=a_0\nabla_{\bf n}(\ell_{\rm b}+\ell)\quad,
\label{eq:slirat}
\end{align} 
where $\ell$ denotes the arc-length of a filament corresponding to angle $\phi$ at center-line distance $s$, with
\begin{align}
\ell(s,\phi)= \int_0^s|\partial_s{\bf R}({s}',\phi)|\d {s}'\quad.\label{eq:len}
\end{align}
The length offset at the base defined as the mismatch between the center line and filament $\phi$ is denoted $\ell_{\rm b}(\phi)$. Correspondingly, the sliding displacement at the base is $\Delta_{\rm b}=a_0\nabla_{\bf n}\ell_{\rm b}$. Here, the normal derivative is defined by  $\nabla_{\bf n}f=(n^s/|{\bf R}_s|)\partial_s f+(n^\phi/|{\bf R}_\phi|)\partial_\phi f$, where $n^s$ and $n^\phi$ are the components of ${\bf n}=n^s{\bf e}_s+n^\phi{\bf e}_\phi$. The filaments splay is defined as 
\begin{align}
\Gamma={\bf e}_a\cdotp\partial_\phi{\bf e}_s\quad,
\label{eq:splay}
\end{align} 
and corresponds to the out of plane rotation of the filament tangent vector when changing $\phi$.

Molecular motors can induce sliding displacement and filament splay by generating active stress conjugate to these strains, see Fig.~\ref{fig:rates}B. These conjugate stresses are the motor force $f_{\rm m}$, which tends to slide filaments apart, and the motor torque $m_{\rm m}$ which tends to induce splay. 

The geometry of the filaments is fully characterized by the {axonemal curvatures}, $\Omega={\bf e}_3\cdotp\partial_s{\bf e}_2$ and $\Theta =-{\bf e}_3\cdotp\partial_s{\bf e}_1$, the twist $\Pi={\bf e}_2\cdotp\partial_s{\bf e}_1$, and the length offset $\ell_{\rm b}(\phi)$. To linear order in the curvatures, we can express the sliding displacement and the splay as
\begin{align}
\Delta(s,\phi)&\approx \Delta_{\rm b}(\phi)-a_0^2\Pi(s)+a_0\cos(\phi)\int_0^s\Omega(s')\d s'
\nonumber\\
&+a_0\sin(\phi)\int_0^s\Theta (s')\d s'\quad,\nonumber\\
\Gamma(s)&\approx a_0\Pi(s)\quad.\label{eq:sliding}
\end{align}
An analogous expansion can also be made for $C_1$ and $C_2$, see Appendix. Note that, to lowest order in the deformations, the splay directly corresponds to axonemal twist.

\section{Work functional} The mechanical properties of the axoneme are characterized by the elasticity of filaments and linkers, and the active forces and torques generated by molecular motors. We introduce the work functional $G$ that describes the mechanical work performed to induce axonemal deformations:
\begin{align}
G &=\int_0^L\int_0^{2\pi}\bigg\{  \frac{\kappa_1}{2}C_1^2 +\frac{\kappa_2}{2}C_2^2 +\frac{k_{\rm s}}{2}\Delta^2 +\frac{k_{\rm r}}{2}(\ell_{\rm b}+\ell-s)^2+f_{\rm m}\Delta\nonumber\\
& +m_{\rm m}\Gamma+\sigma_\phi a_0^2(\partial_\phi\varphi-1)+\sigma_aa_0(a-a_0)+\frac{\Lambda}{2}|\partial_s{\bf r}|^2\bigg\}\d s\d\phi\nonumber\\
&+\int_0^{2\pi}\left\{\frac{K_{\rm s}}{2}\Delta_{\rm b}^2 +\frac{K_{\rm r}}{2} \ell_{\rm b}^2 \right\}\d \phi\quad.\label{eq:energy}
\end{align}
Here $\kappa_1$ and $\kappa_2$ are bending {rigidities} corresponding to deformation of the filaments tangent and perpendicular to the cylindrical sheet. The sliding stiffness of elements that link neighboring doublets is denoted $k_{\rm s}$. Similarly  $k_{\rm r}$ denotes the radial stiffness of sliding linkers between filaments and the central pair. { These elastic constants relate to those of an individual filament by a geometric factor of $9/2\pi$, for example $\kappa_1=9\kappa_{1,{\rm db}}/2\pi$, with $\kappa_{1,{\rm db}}$ the doublet bending rigidity.} The tangential stress $\sigma_\phi$  and the radial stress $\sigma_a$ are Lagrange multipliers to impose the constraints $\varphi=\phi$ and $a= a_0$. The Lagrange multiplier $\Lambda$ imposes the constraint $|\partial_s{\bf r}|=1$. Finally,  $K_{\rm s}$ and $K_{\rm r}$ are basal stiffnesses between neighboring filaments and between filaments and the central pair, respectively. The mechanical work performed by motors is given by $f_{\rm m}\Delta$  and $m_{\rm m}\Gamma$.

In the following, Fourier modes in $\phi$ will play an important role. Therefore we express the $\phi$ dependence of variables in Eq.~\ref{eq:energy} by Fourier series. For instance, the motor force and sliding displacement can be written as
\begin{align}
f_{\rm m} &= f_{{\rm m}}^{(0)} +f_{{\rm m}}^{(1)}\cos(\phi)+f_{{\rm m}}^{(2)}\sin(\phi)+\dots\nonumber\\
\Delta &= \Delta^{(0)} +\Delta^{(1)}\cos(\phi)+\Delta^{(2)}\sin(\phi)+\dots \quad,
\label{eq:modesforce}
\end{align}
In this analysis we truncate the series after the first mode for simplicity. Higher modes can be systematically taken into account if they become relevant. Correspondingly, we also expand $f$, $\sigma_\phi$ and $\sigma_a$. Upper indices in parenthesis in the following always denote azimuthal Fourier modes.

\section{Radial and transverse stress in the axoneme}
A bent and twisted axoneme exhibits radial stress $\sigma_a$ and transverse stress $\sigma_{\phi}$,  which can be key in regulating the motor activity. To determine their values, we first establish a force balance for the twist of the axoneme. In the case where twist relaxes quickly, the corresponding force balance is quasi-static. The torque balance $\delta G/\delta\Pi=0$ then implies
\begin{align}
a_0^4k_{\rm s}\Pi -a_0^2\kappa_1\ds^2{\Pi}=a_0m_{{\rm m}}-a_0^2{f}_{{\rm m}}^{(0)}\quad.\label{eq:twist}
\end{align}
This equation shows that the cilium is  twisted by motor torques $m_{\rm m}$ and also by the zeroth harmonic of motor force $f_{{\rm m}}^{(0)}$. The corresponding twist stiffness is provided by the doublet sliding stiffness $k_{\rm s}$. Since twisting the cilium involves bending of the doublets, their bending stiffness $\kappa_1$ couples to twist. The competition of sliding and bending of the filaments during twist is  characterized by the length-scale $d=\sqrt{\kappa_1/a_0^2k_{\rm s}}$, at which twist deformations decay along the axoneme in response to spatially localized motor forces or torques.

\begin{figure}[t]
\centering
\includegraphics{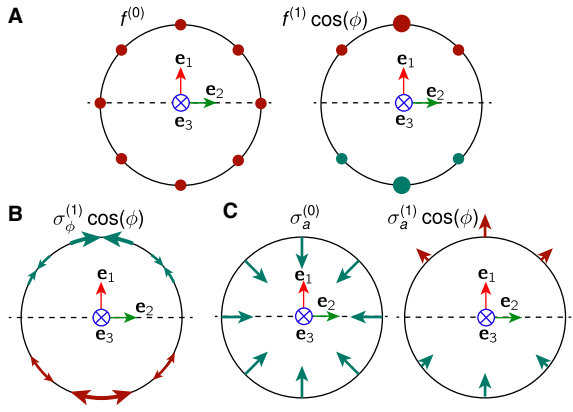}
\caption{\label{fig:transversechiral}{\bf Azimuthal stress profile.}  {\bf A}. Components of the sliding force. The zeroth mode is homogeneous and creates twist. The first mode, positive above the dashed line (beating plane) and negative below, is responsible for in-plane bending. {\bf B} and {\bf C}. The transverse stress has a first mode, and the radial stress has a zeroth and also a first mode. These first modes, possible only in the presence of twist, can regulate the first harmonic of the sliding force.}
\end{figure}

The transverse  and radial stresses can be calculated from the force balances $\delta G/\delta\varphi=0$ and $\delta G/\delta a=0$ as Lagrange multipliers to impose the constraints $\varphi=\phi$ and $a=a_0$. The full expressions of $\sigma_a$ and $\sigma_\phi$ are given in the Appendix. In the simple case of almost planar deformations with  $\Theta =0$ and $\Pi$ small we have
\begin{align}
\sigma_{\phi}^{(1)}&=M_{{\rm m}}\Omega/a_0^2\label{eq:trans}\\
\sigma_a^{(0)}&=\frac{(F^{(1)}+ F_{{\rm r}}^{(2)})\Omega}{2a_0}\label{eq:zeroth}
\\
\sigma_a^{(1)}&=2f^{(1)}\Pi+3\kappa_1\Omega\ds \Pi/a_0+\Omega M_{{\rm m}}\quad,
\label{eq:fn2}
\end{align}
where we have introduced the integrated torque  $M_{\rm m}=-\int_s^Lm_{\rm m}\d s'$ and the integrated net sliding force $F^{(1)}=-\int_s^Lf^{(1)} \d s'$, with $f^{(1)}=f_{{\rm m}}^{(1)}+k_{\rm s}\Delta^{(1)}$, as well as the radial force $F_{{\rm r}}^{(1)}=-\int_s^Lk_{\rm r}(\ell_{\rm b,1}+\ell_1-s) \d s'$. Note that $\sigma_{\phi}^{(2)}$ and $\sigma_{\phi}^{(0)}$ vanish, while the first angular mode $\sigma_{\phi}^{(1)}$ of the transverse stress is proportional to curvature $\Omega$. The zeroth mode of the radial stress, in Eq.~\ref{eq:zeroth}, is a generalization of the normal stress in two dimensional models \cite{lindemann1994model,mukundan2014motor,bayly2014equations}. Figure~\ref{fig:transversechiral} depicts the angular profiles of radial and transverse stresses.

\section{Axonemal dynamics}
The dynamics of the axoneme is governed by a balance of fluid friction forces and axonemal forces
\begin{align}
\partial_t{\bf r}=-\left(\xi_\perp^{-1}({\bf e}_1{\bf e}_1+{\bf e}_2{\bf e}_2)+\xi_{\parallel}^{-1}{\bf e}_3{\bf e}_3\right)\cdotp\frac{\delta G}{\delta {\bf r}}\quad.\label{eq:eom}
\end{align}
Here $\delta G/\delta{\bf r}$ is the axonemal force, and $\xi_\perp$ and $\xi_\parallel$ are the friction coefficients per unit length perpendicular and tangential to the axonemal axis. The explicit expression for $\delta G/\delta{\bf r}$ is given in the Appendix. From Eq.~\ref{eq:eom} we can obtain dynamic equations for the axonemal curvatures $\Omega$ and $\Theta $ given by $\partial_t\Omega=-{\bf e}_2\cdotp\partial_t\ds^2{{\bf r}}$ and $\partial_t\Theta ={\bf e}_1\cdotp\partial_t\ds^2{{\bf r}}$.

In the following we will focus on the case of almost planar and weakly twisted axonemal shapes. Physically this corresponds to enforcing the constraint $\Theta =0$. { The non-linear equation of motion of $\Omega$ is to linear order in $\Pi$ given by
\begin{align}
\xi_\perp\partial_t\Omega&=-\bar{\kappa}\ds^4{\Omega}+\bar{a}\ds^3{f}^{(1)}\label{eq:plane}\\
&+\partial_s^2(\Omega\tau)+(\xi_\perp/\xi_\parallel)\partial_s((\Omega^2(\bar{\kappa}\ds{\Omega}-\bar{a}f^{(1)})+\Omega\ds{\tau}))\nonumber
\end{align}
where $\bar{\kappa}=\pi(\kappa_1+\kappa_2)$ is an effective bending rigidity and $\bar{a}=\pi a_0$. The center-line tension $\tau$ is related to the Lagrange multiplier $\Lambda$ introduced in Eq.~\ref{eq:energy} \cite{camalet_generic_2000}, see Appendix.}

The dynamic shape equation \ref{eq:plane} for the curvature $\Omega$ is a generalization of the previously introduced shape equation for two dimensional beats \cite{machin_1958, brokaw_bend_1971, camalet_generic_2000}. Note that Eq.~\ref{eq:plane} not only describes beats in a plane, but for $\Pi\neq0$ it describes 
 beats on a twisted two-dimensional manifold. The resulting three dimensional shapes  ${\bf r}(s)$ 
 can be determined from  $\Omega$ and $\Pi$, which are solutions to Eqns.~\ref{eq:plane} and 
 \ref{eq:twist}, by integrating  $\ds{\bf e}_3=-\Omega{\bf e}_2$ and $\ds{\bf e}_2 =\Omega{\bf e}_3-\Pi{\bf e}_1(s=0)$. Thus, although we constrained $\Theta =0$, the twist causes out of plane bending.  To impose the constraint of $\Theta =0$ in the presence of twist, the component $f^{(2)}$ of the sliding force 
 becomes a Lagrange multiplier that corresponds to the force introduced by structural elements.

\section{{Self-organized beating by motor control feedback}}
{We now focus our attention on self-organized beating patterns with small amplitude waveforms and angular beat frequency $\omega$. In this case, the periodic flagellar beat is powered by oscillating sliding forces, which we write in frequency representation as 
\begin{align}
f&=\tilde{f}_0+\tilde{f}_1\e^{i\omega t}+\tilde{f}_{-1}\e^{-i\omega t}+\ldots
\end{align}
where $\tilde{f}_0$ is time independent, $\tilde{f}_1=\tilde{f}_{-1}^*$ is the amplitude of the fundamental Fourier mode, and higher frequency harmonics have been omitted for simplicity. We also define time Fourier modes of the azimuthal force components denoted $\tilde{f}_k^{(n)}$, as well as of the transverse stress $\tilde{\sigma}_{\phi,k}^{(n)}$, sliding $\tilde{\Delta}_{k}^{(n)}$, and torque $\tilde{m}_{{\rm m},k}$,  where $k=-1,0,1$ and $n=0,1,2$.

\begin{figure*}[t]
\centering
\includegraphics{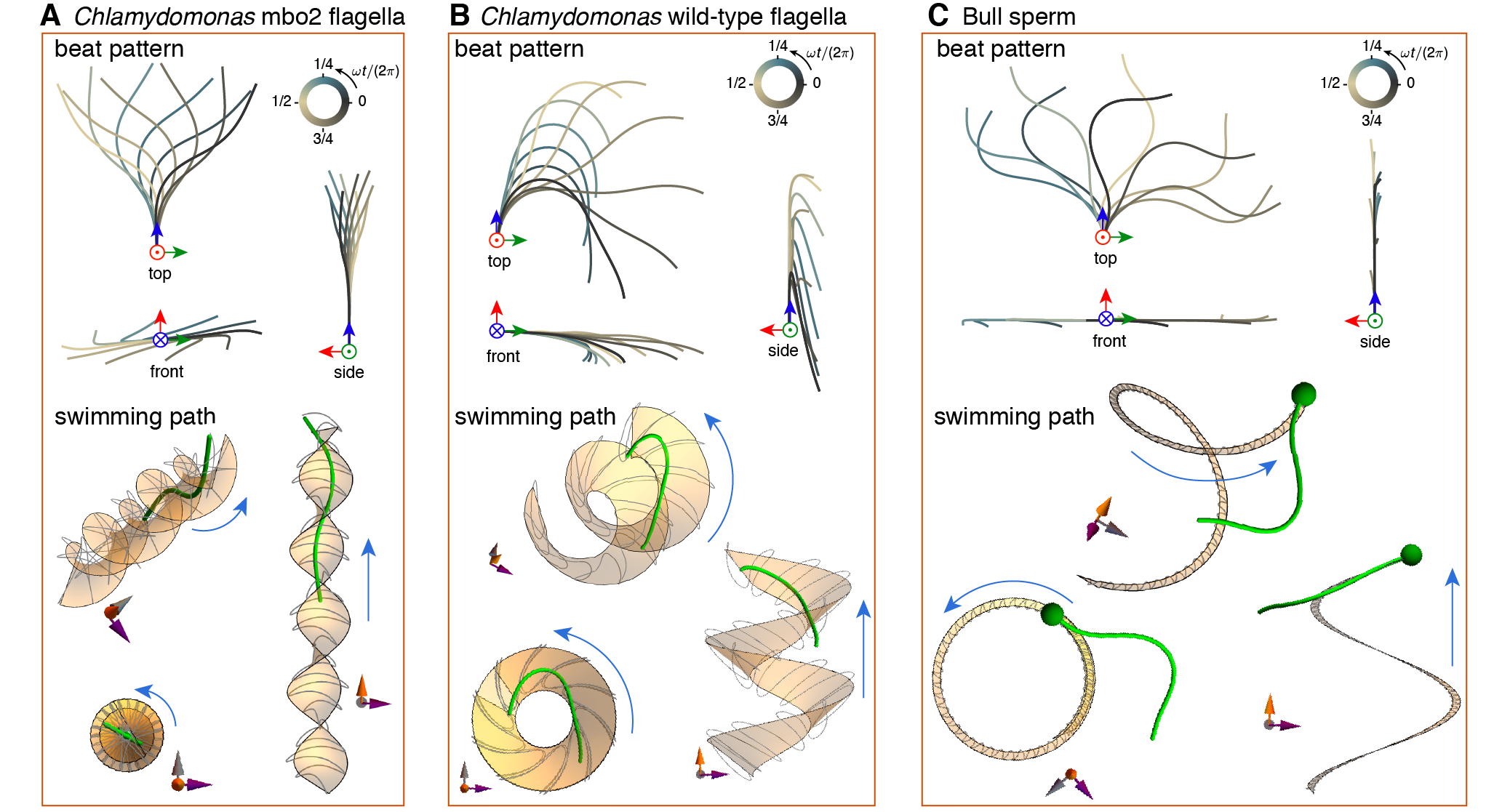}
\caption{\label{fig:beats}{\bf Twisted beating patterns and swimming trajectories.} {\bf A}, {\bf B} and {\bf C}. Above, top, side, and front view of self-organized beating pattern regulated by transverse stress (A and B) or sliding (C). Due to the twist, all beating patterns are non-planar, however only the waveforms in B and C are asymmetric. Below, three different views of the swimming path with a lab-frame for reference. The gray line shows the path of the basal point, the yellow twisted surface the average trajectory, and the blue arrow indicates the swimming direction. For the symmetric beating pattern (A) the swimming surface forms a twisted ribbon, while for the asymmetric cases it forms a helix (B and C). Note that due to the presence of the head the precession of the basal end in C is much smaller than in A and B.}
\end{figure*}

The motor forces and torques are the result of a feedback regulation of motors by axonemal deformations \cite{bell_models_1978, julicher_spontaneous_1997}. Here we focus on the case in which the oscillating instability occurs via the oscillating force amplitude $\tilde{f}_1^{(1)}$, but not the oscillating torque amplitude $\tilde{m}_{{\rm m},1}$. We propose that motor regulation occurs through the sensitivity of the motor function to local axonemal deformations and stresses. For simplicity we illustrate our ideas by focusing on regulation by sliding displacement \cite{brokaw_molecular_1975,camalet_generic_2000, hilfinger2008chirality} and by molecular deformations induced by the transversal stress \cite{lindemann1994model, mukundan2014motor, bayly2014equations}. We therefore write the oscillating motor force to linear order as
\begin{align}
\tilde{f}_1^{(1)}=\chi (\omega)\tilde{\Delta}_1^{(1)}+\zeta(\omega)\tilde{\sigma}_{\phi,1}^{(1)}\quad,\label{eq:mot}
\end{align}
where $\chi(\omega)$ and $\zeta(\omega)$ are complex frequency-dependent 
linear response coefficients describing effects of sliding displacement and transverse stress, respectively \cite{camalet_generic_2000}.} 

\section{{Symmetric and asymmetric twisted beating patterns}}
We now discuss beating patterns that arise by an oscillating instability from an initially
nonoscillating state via a Hopf bifurcation. This instability can be driven by the mechanical feedbacks
mediated by the regulation of motors by sliding displacements or tangential stresses described by
Eq (\ref{eq:mot}). The nonoscillating state is characterized by a time independent curvature 
$\Omega=\tilde{\Omega}_{0}$ and a time independent twist $\Pi=\tilde \Pi_{0}$. We again consider the simple
case $\Theta =0$. Starting from this non-oscillating state, an oscillating mode that represents the flagellar beat emerges at the Hopf bifurcation. This mode can be characterized by the Fourier amplitude of the fundamental frequency component $\tilde{\Omega}_{1}$, which becomes non-zero beyond the bifurcation point.

{
We now discuss the symmetry properties of the emerging beating patterns. We first consider the simple case of vanishing static twist $\Pi=0$, for which beats are confined to a plane. In this case we can distinguish symmetric beats with vanishing average curvature $\tilde\Omega_0=0$ and asymmetric beats. Symmetric beats are mirror symmetric within the beat plane \cite{camalet_generic_2000} and the swimming trajectories are straight lines within the plane. This mirror symmetry is broken in the asymmetric case, for which swimming trajectories are circles in the plane \cite{vkthesis, friedrich2010high}. Generally, the static twist does not vanish. The beating pattern then is confined to a twisted two dimensional manifold, see Fig.~\ref{fig:beats}, bottom. Again we can distinguish symmetric and asymmetric twisted beats. In the symmetric case with $\tilde\Omega_0=0$ the beat is now symmetric with respect to $\pi$ rotations with a rotation axis tangential to the manifold, see Fig.~\ref{fig:beats}A, and swimming trajectories are straight lines on the manifold. This symmetry is broken in the case of asymmetric twisted beats, which exhibit helical swimming paths, see Fig.~\ref{fig:beats}B.

The static twist is determined from Eq.~\ref{eq:twist}.  The static curvature of asymmetric beats follows from the force balance $\bar{\kappa}\tilde{\Omega}_{0}=\bar{a}\tilde F^{(1)}_0$. For simplicity we illustrate the main ideas where the static force and torques act at the distal end, in which  case  $F^{(1)}_0$ and $M_{{\rm m},0}$ are $s$ independent. As a consequence the static curvature is constant and the average shape is a twisted circular arc.  } 
We can now express the dynamic equation for the beat shape. For the case of a symmetric twisted beat, we find
\begin{align}
i\omega\xi_\perp\tilde{\Omega}_{1}=-\bar{\kappa}\ds^4{\tilde{\Omega}}_{1}+\pi{a}_0^2\chi\ds^2{\tilde{\Omega}}_{1}+\pi a_0\beta\ds^3{\tilde{\Omega}}_{1}\quad.\label{eq:crit}
\end{align}
Here  $\beta(\omega)=(M_{{\rm m},0}/a_0^2)\zeta(\omega)$ plays the role of an effective motor control feedback by curvature. The term proportional to $\chi$ describes a motor control feedback by sliding displacements.  Solving Eq.~\ref{eq:crit} provides the time dependence of the curvature $\Omega$ from which we determine the axonemal shape by integration along the arc-length. An example of such a beat is shown in  Fig.~\ref{fig:beats}A in the case where oscillations are generated by motors that are regulated via transverse stresses. The resulting waveform is non-planar and the swimming path corresponds to a twisted ribbon. This beating patterns is similar to that of the {\it Chlamydomonas} mutant {\it mbo2} \cite{vkthesis, segal1984mutant}. 

Asymmetric beating patterns can be studied in a similar way, by expanding Eq.~\ref{eq:plane} near a state with constant static curvature $\tilde{\Omega}_{0}$. This leads to a  generalization of Eq.~\ref{eq:crit} describing asymmetric beats, see Appendix. An example of such a beating pattern for the case of motor regulation by transverse stresses is shown in Fig.~\ref{fig:beats}B. This beating pattern is non-planar and asymmetric, and the resulting swimming path is a helical ribbon. This waveform is similar to the one observed for the wild-type {\it Chlamydomonas} axoneme \cite{eshel1987new, vkthesis}.

Alternatively we can also discuss beats by motor regulation via sliding. It has been suggested that sliding control likely governs the beat shape of bull sperm \cite{riedelkruse_how_2007}. In Fig.~\ref{fig:beats}C we show such a beating pattern of asymmetrically beating axoneme with motors regulated via sliding. The result is a waveform similar to that of a freely swimming bull sperm. Due to the friction of the sperm head (green sphere in Fig.~\ref{fig:beats}C) the resulting helical ribbon is narrow as compared to Fig.~\ref{fig:beats}B, see also \cite{jikeli2015sperm,su2012high}.

\section{Discussion}
The role of the three dimensional architecture of the axoneme for motor regulation and beat generation is poorly understood. Here, using a three dimensional continuum mechanical model of the axoneme, we showed that both shear stresses (associated with sliding forces) and transverse stresses (associated with torques) can be used to regulate motors and generate periodic beating patterns. In particular we showed that in the presence of twist transverse stresses are proportional to curvature, and motor regulation by transverse stresses effectively results in control motor activity by axonemal curvature.

Dynein generated shear forces induce relative sliding of microtubule doublets. We show here that motor torques induce splay deformations of microtubule doublets, see Fig.~\ref{fig:rates}. Because microtubule splay deformations lead to axonemal twist, we conclude that motor torques twist the axoneme even in the absence of shear forces. As a consequence of the twist beating patterns are in general chiral.

Axonemal twist is governed by torsional stiffness. Our work shows that sliding elastic linkers between neighboring microtubule doublets provide an effective torsional stiffness $a_0^4 k_{\rm s}$ of the axoneme, see Eq~\ref{eq:twist}. In addition there could be a contribution to the torsional stiffness of the axoneme from the torsional stiffness $\kappa_3$ of individual microtubule doublets, which for simplicity we have not included in our discussion. This contribution becomes relevant when doublets are constrained not to rotate around their axis, and results in a net torsional stiffness $\kappa_3+a_0^4 k_{\rm s}$. Estimating $\kappa_3\sim\kappa_1$, we have that $\kappa_3/(a_0^4 k_{\rm s})\sim(d/a_0)^2\sim10^3$. In this case the  torque necessary to twist the axoneme would be larger, and the twist would decay faster along the axonemal length, see Appendix.

Our work shows that feedback control of motors both by shear or by transverse stress can induce time periodic bending waves in three dimensions via a dynamic instability. We show that motor regulation by transverse  stresses requires motor torques that twist the axoneme. This mechanism effectively gives rise to motor control by axonemal curvature, which is very reliable and even works for short axonemes for which sliding control does not induce traveling waves \cite{fittingarx}.  These waveforms are three dimensional because of the chirality of the axoneme and the resulting axonemal twist, see Fig.~\ref{fig:beats}. These beating patterns in general generate swimming trajectories that are either helical paths or can be represented as twisted ribbons.  Such swimming paths have indeed been observed experimentally for different sperm cells \cite{su2013sperm, su2012high, jikeli2015sperm}. If such swimmers are observed near surfaces the chirality of the beat leads to circular trajectories  \cite{elgeti2010hydrodynamics} which have been reported for various systems \cite{bessen1980calcium,friedrich2010high}.

The beating patterns of {\it Chlamydomonas} and of sperm often exhibit periodic deformations around an average curvature. In our work we introduced this average curvature via an asymmetry of static forces at the distal end. However the molecular origin of these beat asymmetries remains unclear. We think that our continuum mechanical model of the axoneme will be an important tool to understand the mechanical origin of beat asymmetries and the selection of the beat plane of flagellar beats.  

\section{Appendix}
\subsection{Curvatures of filaments for small deformations}
\label{sec:expand}
The filaments on the cylindrical surface have curvatures  $C_1(s,\phi)$ and $C_2(s,\phi)$. These are functions of the principal curvatures of the center-line, $\Omega(s)$ and $\Theta (s)$, as well as the twist, $\Pi(s)$. For the case of small deformations, we have
\begin{align}
C_1(s,\phi)&\approx\Omega(s)\cos(\phi)+\Theta (s)\sin(\phi)-a_0{\partial_s \Pi}\nonumber\\
C_2(s,\phi)&\approx\Theta (s)\cos(\phi)-\Omega(s)\sin(\phi)\quad,\nonumber
\end{align}
where only terms linear in the axoneme curvature are kept, see also \cite{heussinger2010statics}. Note that the rate of twist, and not the twist, affects the curvature on the tangent plane $C_1$. This is because, as shown in Eq.~\ref{eq:sliding}, a constant twist corresponds to a constant sliding. Constant twist contributes as a higher order correction to $C_2$. 

\subsection{Radial and transverse stress}
\label{sec:radtran}
To obtain the radial stress $\sigma_a$ and the transverse stress $\sigma_\phi$ we use the force balances given by $\delta G/\delta a=0$ and $\delta G/\delta \varphi=0$. The procedure is straightforward, see \cite{camalet_generic_2000,hilfinger2008chirality,mukundan2014motor} for similar calculations, and for $\sigma_a$ directly gives
\begin{align}
\sigma_a&=2(f-m_{\rm m}/a_0)\Pi+F(\Omega\cos(\phi)+\Theta\sin(\phi) )/a_0\nonumber\\
&+F_{\rm r}(\Omega\sin(\phi)-\Theta\cos(\phi) )/a_0\nonumber\\
&-\kappa_1((\partial_s\Pi)^2-2\partial_s\Pi(\Omega\cos(\phi)+\Theta \sin(\phi))/a_0)\quad.\nonumber
\end{align}
For the case $\Theta =0$ and to linear order in $\Pi$, the zeroth and first azimuthal modes are those in Eqs.~\ref{eq:zeroth} and \ref{eq:fn2}, where we have used the integral of Eq.~\ref{eq:twist}. In the case of the transverse stress, the force balance   results in 
\begin{align}
a_0^2\partial_\phi\sigma_\phi&={\kappa_1}(\Omega\cos(\phi)+\Theta \sin(\phi)-a_0\ds\Pi)(\Theta \cos(\phi)-\Omega\sin(\phi)) \nonumber\\
&-{\kappa_2}(\Theta \cos(\phi)-\Omega\sin(\phi))(\Theta \sin(\phi)+\Omega\cos(\phi)) \nonumber\\
& -a_0F(\Theta\cos(\phi) -\Omega\sin(\phi))-a_0F_{\rm r}(\Omega\cos(\phi)+\Theta\sin(\phi) ).\nonumber
\end{align}
For the case $\Theta =0$ we have $a_0^2\partial_\phi\sigma_\phi=(a_0F^{(0)}+a_0\kappa_1\ds\Pi)\Omega\sin(\phi)+\ldots$. Using Eq.~\ref{eq:twist}  and integrating gives the harmonic in Eq.~\ref{eq:trans}.

\subsection{General equations of motion} 
In the case in which the twist $\Pi$ relaxes fast,  it is calculated from Eq.~\ref{eq:twist}, see \cite{hilfinger2008chirality} for  a characterization of twist dynamics. The force balance describing the dynamics of the center-line $\vec{\bf r}$ is given in Eq.~\ref{eq:eom}, where
\label{sec:planar}
\begin{align}
\frac{\delta G}{\delta \vec{r}}&=-\partial_s\Big\{\hat{\bf e}_1\left(-\bar{a}(F_{\rm t}^{(1)}\Pi+f^{(2)})+\bar{\kappa}\Omega\Pi+\bar{\kappa}\ds{\Theta}\right)\nonumber\\
&+\hat{\bf e}_2\left(-\bar{a}(F_{\rm t}^{(2)}\Pi-f^{(1)})+\bar{\kappa}\Theta \Pi{-}\bar{\kappa}\ds{\Omega}-\hat{\bf e}_3\tau\right)\Big\},\nonumber
\end{align}
where we have introduced the total force modes $F_{\rm t}^{(1)}=F^{(1)}+F_{\rm r}^{(2)}$ and $F_{\rm t}^{(2)}=F^{(2)}+F_{\rm r}^{(1)}$, and the tension $\tau=2\pi\Lambda-\bar{a}(F^{(1)}\Omega+F^{(2)}\Theta )-\bar{\kappa}(\Omega^2+\Theta ^2)$. Taking the time derivative of the constraint $\ds\vec{{\bf r\,}}^2=1$ we arrive at $\hat{\bf e}_3\cdotp\ds\dt\vec{\bf r} =0$, which together with Eq.~\ref{eq:eom} provides the equation for the tension. Finally, to calculate the modes of the basal length $\ell_{\rm b}^{(n)}$ we use the sliding force balances $\delta G/\delta \ell_{\rm b}^{(n)}=0$. For $n=1$ and $n=2$ the equations are $K_{\rm t}\ell_{\rm b}^{(1)} = F^{(2)}_{\rm t}(0)$ and $K_{\rm t}\ell_{\rm b}^{(2)} =F^{(1)}_{\rm t}(0)$, while doing $\delta G/\delta \ell_{{\rm b}}^{(0)}=0$ results in $\ell_{{\rm b}}^{(0)}=0$. Here, we have defined the total basal stiffness $K_{\rm t}=K_{\rm s}+K_{\rm r}$.

The boundary forces and torques exerted by the filament correspond to the boundary terms of  $\delta G/\delta \vec{r}$ and $\delta G/\delta \Pi$, see \cite{camalet_generic_2000, hilfinger2008chirality, mukundan2014motor}. Balancing these by external forces $\vec{F}_{\rm ext}$ and torques $\vec{T}_{\rm ext}$ we have at the base $s=0$:
\begin{align}
\vec{F}_{\rm ext}&=-\Big\{\hat{\bf e}_1\left(-\bar{a}(F_{\rm t}^{(1)}\Pi+f^{(2)})+\bar{\kappa}\Omega\Pi+\bar{\kappa}\ds{\Theta}\right)\nonumber\\
&+\hat{\bf e}_2\left(-\bar{a}(F_{\rm t}^{(2)}\Pi-f^{(1)})+\bar{\kappa}\Theta \Pi-\bar{\kappa}\ds{\Omega}\right)-\hat{\bf e}_3\tau\Big\}\nonumber\\
{ \vec{T}_{\rm ext}}&=(\bar{a} F^{(2)}-\bar{\kappa}\Theta -\pi\kappa_1a_0^2\Omega\ds\Pi)\hat{e}_1\nonumber\\
&+(-\bar{a} F^{(1)}+\bar{\kappa}\Omega+\pi\kappa_1a_0^2\Theta \ds\Pi)\hat{e}_2\nonumber\\
M_{\rm ext}&=\kappa_1a_0\ds \Pi\nonumber
\quad,
\end{align}
The lack of a third component in the torque balance comes from neglecting the twist dynamics \cite{hilfinger2008chirality}. The third moment balance comes from the contribution of filament bending, and involves an additional external moment $M_{\rm ext}$. At the tip $s=L$ the boundary conditions are analogous. 
In this work we considered two types of boundary conditions. For a freely swimming flagellum the external forces and torques are null. For a flagellum  attached to a head the torques and forces at the base are $\vec{F}_{\rm ext}=\xi_{\rm trans}\vec{v}$ and $\vec{T}_{\rm ext}=\xi_{\rm rot}\vec{\omega}$, where {$\vec{\omega}=(\hat{e}_3\cdotp\dt \hat{e}_2,\hat{e}_1\cdotp\dt \hat{e}_3,\hat{e}_2\cdotp\dt\hat{e}_1)_{0}$} and $\vec{v}=\dt\vec{r}_0$ are the head's rotational and translational velocities, with the subindex $0$ indicating $s=0$, and $\xi_{\rm trans}$ and $\xi_{\rm rot}$ the corresponding friction coefficients of the head.

\subsection{Asymmetric dynamics}  For asymmetric beats we expand Eq.~\ref{eq:plane} around a static shape given by a constant static curvature $\tilde{\Omega}_{0}=\bar{a}\tilde F^{(1)}_0/\bar{\kappa}$. The dynamic mode then obeys 
\begin{align}
i\omega\xi_\perp\tilde{\Omega}_{1}&=-\bar{\kappa}\ds^4{\tilde{\Omega}}_{1}+\bar{a}\ds^3{f}^{(1)}_1+(1+\xi_\perp/\xi_\parallel)\tilde{\Omega}_{0}\ds^2\tau_1\nonumber\\
&+(\xi_\perp/\xi_\parallel)\tilde{\Omega}_{0}^2(\bar{\kappa}\ds^2\tilde{\Omega}_{1}-\bar{a}\ds f^{(1)}_1)\quad,\nonumber
\end{align}
where $\tau_1$ is obtained from expanding the equation for the tension.

\subsection{Twist by tip-accumulated torques}
If the source of twist is a tip accumulated torque, we have that $m_{\rm m}=M\delta(s-L)$. In the case in which $f_{{\rm m}}^{(0)}=0$,  the solution to Eq.~\ref{eq:twist} is using the boundary conditions $\ds\Pi(s=0)=0$ and $a_0\kappa_1\ds\Pi(s=L)=M$ is $\Pi=({Md}/{\kappa_1 a_0})\sinh(s/d)/\cosh(L/2)$. Note that for $L\ll d$ the twist created changes little along the length. Conversely, for  $L\gg d$ the twist quickly decays away from the tip.

\subsection{Swimming trajectories}
The force density exerted by the cilium in the fluid is $\vec{\bf f}_{\rm fl}= \left(\xi_\perp(\hat{\bf e}_1\hat{\bf e}_1+\hat{\bf e}_2\hat{\bf e}_2)+\xi_{\parallel}\hat{\bf e}_3\hat{\bf e}_3\right)\cdotp\partial_t\vec{\bf r}$. Imposing that the sum of all forces and torques in the fluid must vanish we have  $\xi_{\rm trans} \vec{\bf v}+\int_0^L\vec{\bf f}_{\rm fl}\d s=0$ and $\xi_{\rm rot} \vec{ \omega}+\int_0^L\vec{\bf r}\times\vec{\bf f}_{\rm fl}\d s=0$, where the terms outside the integrals come from the head's drag \cite{perrin1934mouvement, friedrich2010high, johnson1979flagellar}. Given a beating pattern we can calculate $\vec{f}_{\rm fl}$ and use these equations to obtain the translational and rotational velocities at each instant during the beat.

\subsection{Parameters used} { The {\it Chlamydomonas} cilium is $L=10.2\,\um$ long and has frequency $\omega/2\pi=73.1\,{\rm Hz}$ \cite{vkthesis}, while {bull sperm} has $L=58.3\,\um$ and $\omega/2\pi=19.8\,{\rm Hz}$  \cite{riedelkruse_how_2007}. For {bull sperm}  {$\tilde{\Omega}_{0}=0.010\,\mu {\rm m}^{-1}$} \cite{riedelkruse_how_2007}, for wild-type {\it Chlamydomonas} we took $\tilde{\Omega}_{0}=0.25\,\mu {\rm m}^{-1}$ and  for {\it mbo2} $\tilde{\Omega}_{0}=0$ \cite{vkthesis, eshel1987new}. The radius we used is $a_0=0.2\,\um$ \cite{pigino2012axonemal}. To  estimate $M$,  note that a density of $~500\,\um^{-1}$ dyneins with a force of $~0.7\,\pN$ accumulated in one distal micron results in a force of $\tilde{F}_0^{(1)}\approx300\,\pN$, compatible with $\tilde{\Omega}_{0}$ for  {\it Chlamydomonas}. If $\sim5\%$ of this force results in a torque over a distance of $0.015\,\um$, we obtain {$M\approx0.25\,\pN\,\um$}, used for {\it Chlamydomonas}. For {bull sperm}, with a smaller $\tilde{\Omega}_{0}$,  we used  {$M\approx0.025\,\pN\,\um$} instead. A doublet has $24$ protofilaments compared to $13$ in a microtubule. The bending stiffness scales as area squared, and so  $\kappa_{\rm db}\approx (24/13)^2\kappa_{\rm mt}\approx80\,{\rm pN}\,\um^2$, with $\kappa_{{\rm mt}}\approx23\,{\rm pN}\,\um^2$ for microtubules \cite{gittes1993flexural}. For simplicity we take $\kappa_{1,{\rm db}}=\kappa_{2,{\rm db}}=\kappa_{{\rm db}}$, which results in $\kappa_1\approx115\,{\rm pN}\,\um^2$ and $\bar{\kappa}=9\kappa_{{\rm db}}\approx700\,{\rm pN}\,\um^2$,  comparable to measurements of {sea urchin sperm} \cite{howard2001mechanics}. The sliding stiffness $k_{\rm s}$ was determined in \cite{pelle2009mechanical}, and corresponds to $d$ between $3.5\,\um$ and $10\,\um$, we chose $d\approx6\,\um$. The resulting distal twist angle for {bull sperm} is {$~0.4\,{\rm rad}$}, in the range of {$~0.25\,{\rm rad}$} obtained in \cite{jikeli2015sperm}. The friction coefficients are $\xi_\parallel\approx2\pi\mu/(\ln(2L/a_0)-1/2)$ and $\xi_\perp\approx2\xi_\parallel$, where $\mu$ is the viscosity \cite{johnson1979flagellar, gray1955propulsion, cox1970motion}. For water  at $22\,{\rm C}$ we have $\mu=0.96\, 10^{-3}{\rm pN\,s}\,\mu{\rm m}^{-2}$, which for $L\approx\,10\,\mu{\rm m}$ results in $\xi_\parallel\approx0.0017\,{\rm pN}\,{\rm s}\,\um^{-2}$. For the head of {bull sperm} we use $\xi_{\rm trans}=6\pi r\alpha_{\rm t}\mu$  and {$\xi_{\rm rot}=8\pi \alpha_{\rm r}r^3\mu$} {\cite{howard2001mechanics}}, where $r$ is the radius of the head and $\alpha_{\rm t}=1/(1-9/16)$ and $\alpha_{\rm r}=1/(1-1/8)$ are  corrections due to the proximity to a wall \cite{leach2009comparison}. In \cite{riedelkruse_how_2007} it was observed that sliding controlled beating patterns require a large head friction. We take $r\approx10\,\um$, large for {bull sperm} but adequate for other species \cite{cummins1985mammalian}. This results in $\xi_{\rm rot}\approx35\,{\rm pN}\,{\rm s}\,\um$ and $\xi_{\rm trans}=0.45\,{\rm pN}\,{\rm s}\,\um^{-1}$. The values of the response coefficients for {\it Chlamydomonas} wild-type beats were $\zeta=-i221\,\um$ and $\chi=-1645\,\pN/\um^2$, for the {\it mbo2} mutant $\zeta=-i235\,\um$ and $\chi=-1916\,\pN/\um^2$, and for bull sperm $\chi=(-5516 -i10138)\,\pN/\um^{2}$.
}


\begin{acknowledgments}
We are thankful to M. Bock and E. M. Friedrich for insightful discussions, and particularly to G. Klindt for suggestions regarding the hydrodynamic aspects of this paper.
\end{acknowledgments}


\end{document}